\documentclass[final,3p,times,twocolumn]{elsarticle}
 \biboptions{comma,sort&compress}
\usepackage{here}
 \usepackage{graphicx}
  \usepackage{epsfig}
\newcommand {\pom} {I\!\!P}
\newcommand {\pomsub} {{\scriptscriptstyle \pom}}

\def\nin{\noindent}

\journal{Nuc. Phys. (Proc. Suppl.)}

\begin{document}

\begin{frontmatter}



\title{%
Diffractive Structure Functions from the H1 and ZEUS Experiments at HERA}

 \author[label1]{Irina A. Korzhavina\corref{cor1}}
  \address[label1]{Experimental High Energy Department, Skobeltsyn Institute of Nuclear Physics, Moscow State University,\\ 
  Vorobjevy Gory 1/2, GSP-1, 119991 Moscow, Russia.}
\cortext[cor1]{On Behalf of the H1 and ZEUS Collaborations} 
\ead{irina@mail.desy.de}

%
\begin{abstract}
\noindent
 The cross section of inclusive diffractive process $ep \to eXp$  
was measured within a wide kinematic range in the H1 and ZEUS 
experiments at HERA. Results obtained by different experiments and 
methods are compatible within measurement uncertainties.  The  
measurements were  subjected to DGLAP 
next-to-leading order QCD  global fits and  the diffractive parton 
distribution functions (DPDFs) of the proton were determined   with 
noticeably reduced uncertainties due to  very high precision of 
the data. The gluon density precision was much improved in fits which 
also included  data on  dijet production in diffractive DIS.
Predictions based on the determined DPDFs are in agreement with the 
measured  inclusive cross-section of diffractive dijet  photoproduction 
and charm production in diffractive DIS at HERA.
The longitudinal diffractive structure 
function $F^D_L$ was measured directly for the first time. 
\end{abstract}



\end{frontmatter}


\section{Introduction}
\nin
Studies of the dissociation of virtual photons, $ \gamma^* p\to Xp $, 
in diffractive deep inelastic $ep$ scattering (DIS), $ep \to eXp$, at 
HERA  contributed significantly to the understanding the dynamics of  
diffraction in terms of quarks and gluons and to further development of 
perturbative Quantum Chromodynamics (pQCD) calculation techniques
\cite{DiffrRevs}.  These processes constitute a large fraction 
($\sim10\%$) of the visible DIS cross section and are mostly 
originating from the hard scattering of partons. Theoretically predicted 
factorisation  \cite{FactorisationTheorem} was  confirmed  in  
measurements of diffractive dijet production by HERA experiments 
H1 and ZEUS. In studies of diffractive scattering it was used for QCD 
fits to measure momentum 
distributions of partons in the proton, called diffractive parton 
distribution functions (DPDFs). 

The differential diffractive DIS cross section  is measured 
in the kinematic variables of the exchanged boson virtuality $Q^2$, 
the inelasticity $y$, the fraction of the  momentum of the proton 
carried by the diffractive exchange $x_{\pomsub}$, the Bjorken 
variable defined for the diffractive exchange $\beta$, and 
the four-momentum transfer squared at the proton vertex $\vert t \vert$.  

The cross section  
may be presented in terms of the diffractive reduced cross section 
$\sigma^{D(4)}_r$ related to the diffractive structure functions, 
$F^{D(4)}_{2/L}$, as: \\
$\sigma^{D(4)}_r(\beta,Q^2,x_{\pomsub},t)$=
$\frac{\beta Q^4}{2\pi\alpha^2Y_+}\frac{d^4\sigma(ep\to eXp)}%
                                       {d\beta dQ^2 dx_{\pomsub} dt}$ \\  
\hspace*{1.5cm}%
$=F^{D(4)}_2(\beta,Q^2,x_{\pomsub},t)-%
                \frac{y^2}{Y_+}F^{D(4)}_L(\beta,Q^2,x_{\pomsub},t)$, \\
where $Y_+=1+(1-y)^2$. 
Like the inclusive structure functions, the  $F^{D(4)}_{2/L}$   are 
defined being  the proton structure functions, probed  
in a process with a fast proton of fractional momentum ($1-x_{\pomsub}$) 
in the final state.

  In accord to the factorisation theorem, diffractive cross section 
can also be presented as a sum of hard scattering cross 
sections for a parton $i$ convoluted with  DPDFs $f^D_i$.
Thus the diffractive structure functions, $F^{D(4)}_{2/L}$ are related 
to the  DPDFs  through the convolutions with coefficient functions 
$C_{2/L,i}$: \\
$F^{D(4)}_{2/L}(\beta,Q^2,x_{\pomsub},t) = \sum_i\int^1_{\beta}%
\frac{dz}{z}C_{2/L,i}(\frac{\beta}{z})f^D_i(z,x_{\pomsub};Q^2,t)$,  \\ 
where z is the longitudinal momentum 
fraction of the parton entering the hard subprocess with respect to the 
diffractive exchange. In the lowest order sub-process  $z = \beta$. 
The inclusion of higher order sub-processes leads to $\beta < z$.

\section{Diffractive Cross Section Measurements}
\nin
At high centre-of-mass energy, diffractive $ep$ scattering 
has two specific signatures.A leading proton  
carrying most ($>90\%$) of the beam energy may be directly 
detected (proton tagging measurement) in the final state 
using Proton Spectrometers (PS: H1 (V)FPS or ZEUS LPS). 
There may also be identified a final state with no hadronic 
activity in a large rapidity region, called Large Rapidity Gap (LRG). 
The LRG separates outgoing (untagged) proton (or (untagged) low mass  
hadronic system $Y$ originating from proton dissociation) from the rest 
of the hadronic system $X$.

\begin{figure}[!hbt] 
\centerline{\includegraphics[width=0.5\textwidth]%
{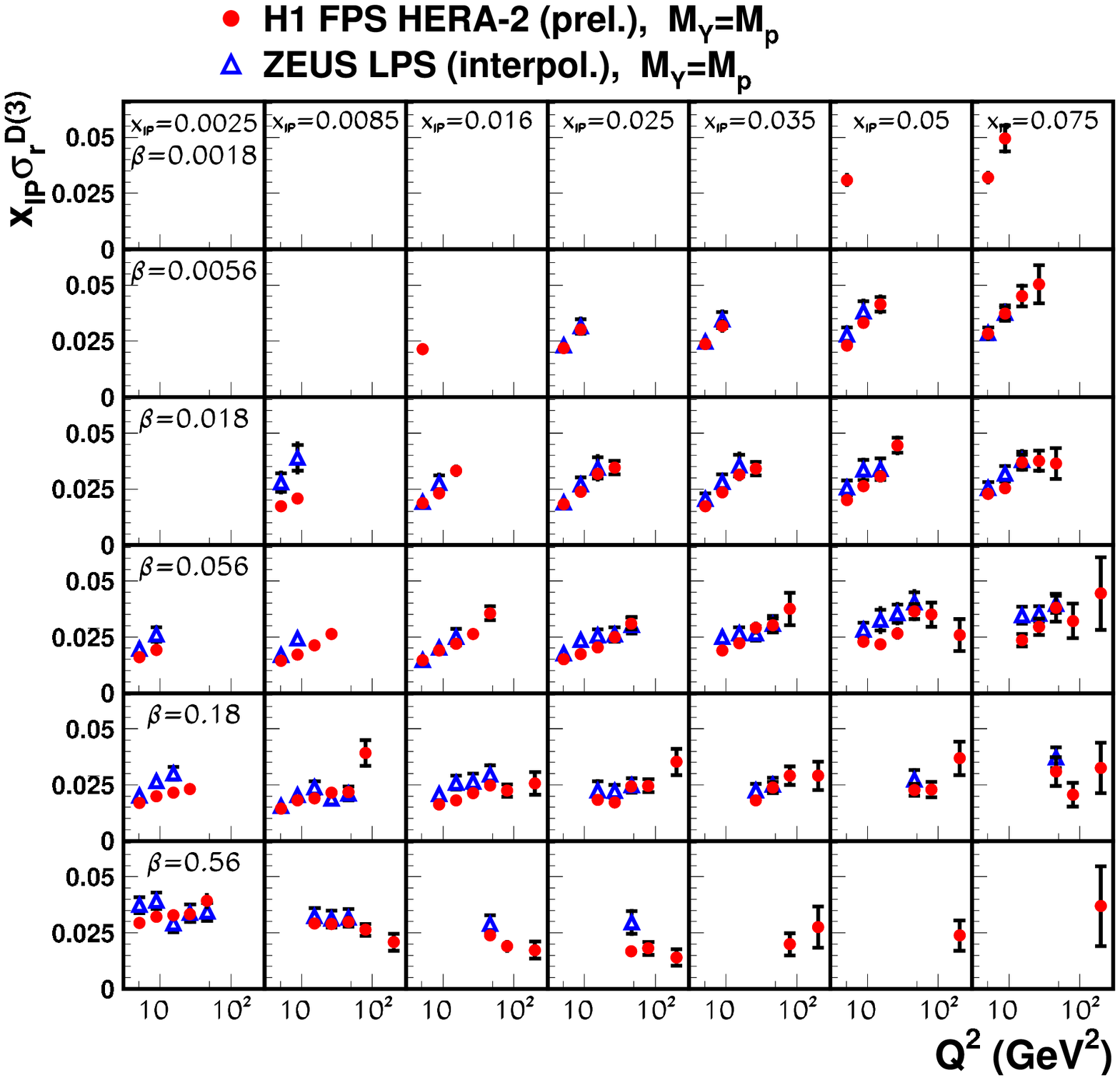}}
\hspace*{0.3cm}
\centerline{\includegraphics[width=0.57\textwidth]%
{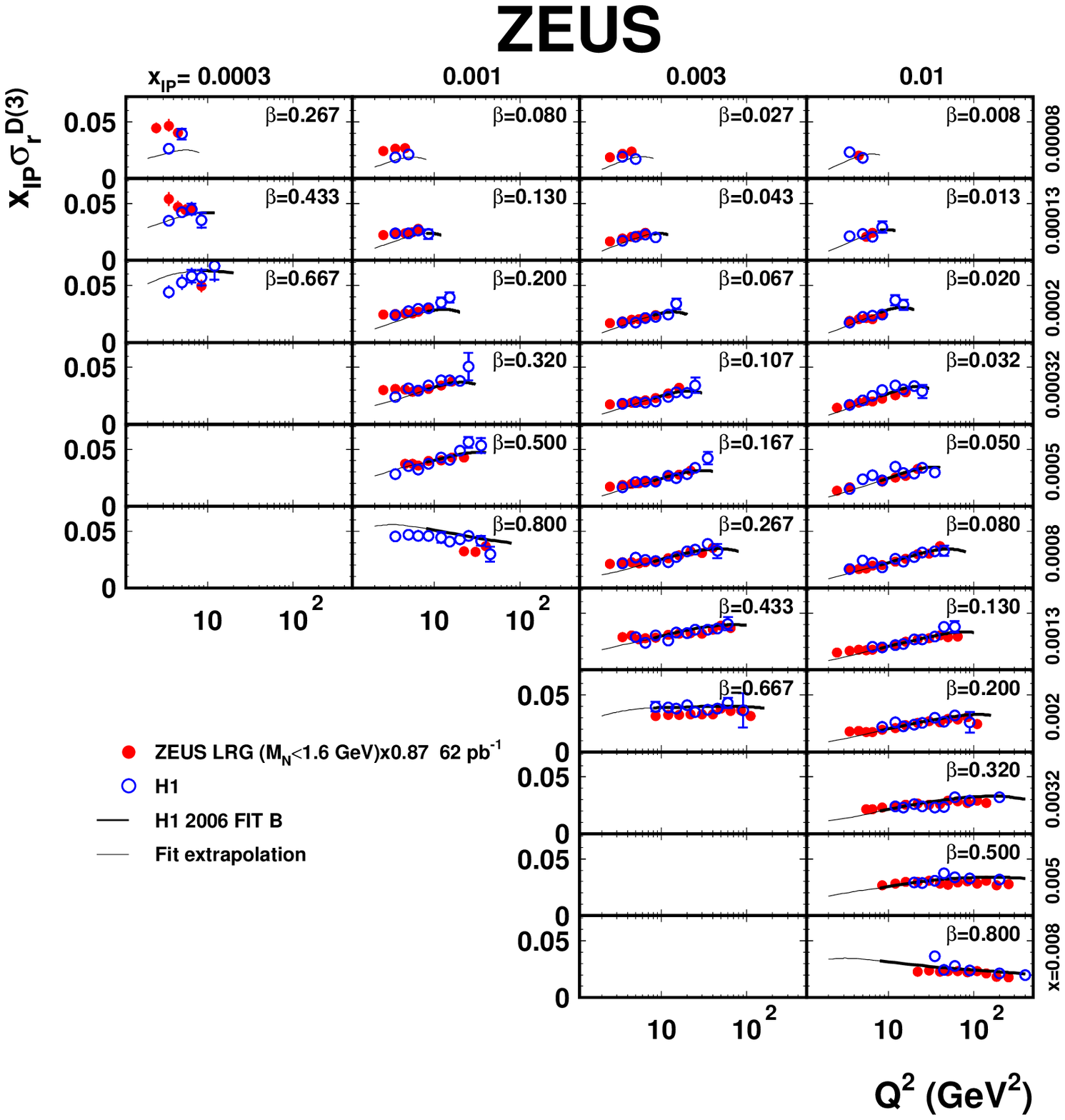}}
\vspace*{-0.3cm}
\caption{%
Comparison between the H1 and ZEUS measurements of the 
diffractive reduced cross section $\sigma^{D(3)}_r$ as a function 
of $Q^2$ in ($x_{\pomsub},\beta$) bins,  
 PS (up) or LRG (down) data being used.
        }
\label{fig:csr3-h1vszeus} 
\end{figure}

Lately  H1 \cite{H1-sigrd} and ZEUS \cite{ZEUS-sigrdlrglps} 
collaborations measured reduced cross sections of the diffractive 
process $ep \to eXY$  with improved precision  compared to earlier 
measurements, both the PS and LRG measurement techniques being used.
Though the ranges accessed by the H1 and ZEUS measurements do not 
quite coincide with each other (Fig.~\ref{fig:csr3-h1vszeus})
both data lie within the kinematic limits: 
 $2<Q^2<700$ GeV$^2$, $\vert t \vert <1.0$ GeV$^2$, 
 $x_{\pomsub}<0.1$, $M_Y(PS)=m_p$  or  $M_Y(LRG)<1.6$ GeV
 and full range of $\beta$.
There is a global normalisation difference at the $\sim 10$--$13\%$ 
level between the measurements of the two experiments 
(Fig.~\ref{fig:csr3-h1vszeus} up) compatible with uncertainties.

To match H1 and ZEUS cross sections the ZEUS data were scaled by the 
factors  consistent with unity accounting for the normalisation 
uncertainties (Fig.~\ref{fig:csr3-h1vszeus} down).
 Apart from the normalisation discrepancy, H1 and ZEUS  high precision 
measurements show good agreement in shapes of distributions
throughout most of the phase space studied  no matter what a 
measurement method used.

\begin{figure}[!hbt] 
\centerline{%
\includegraphics[width=0.17\textwidth]%
{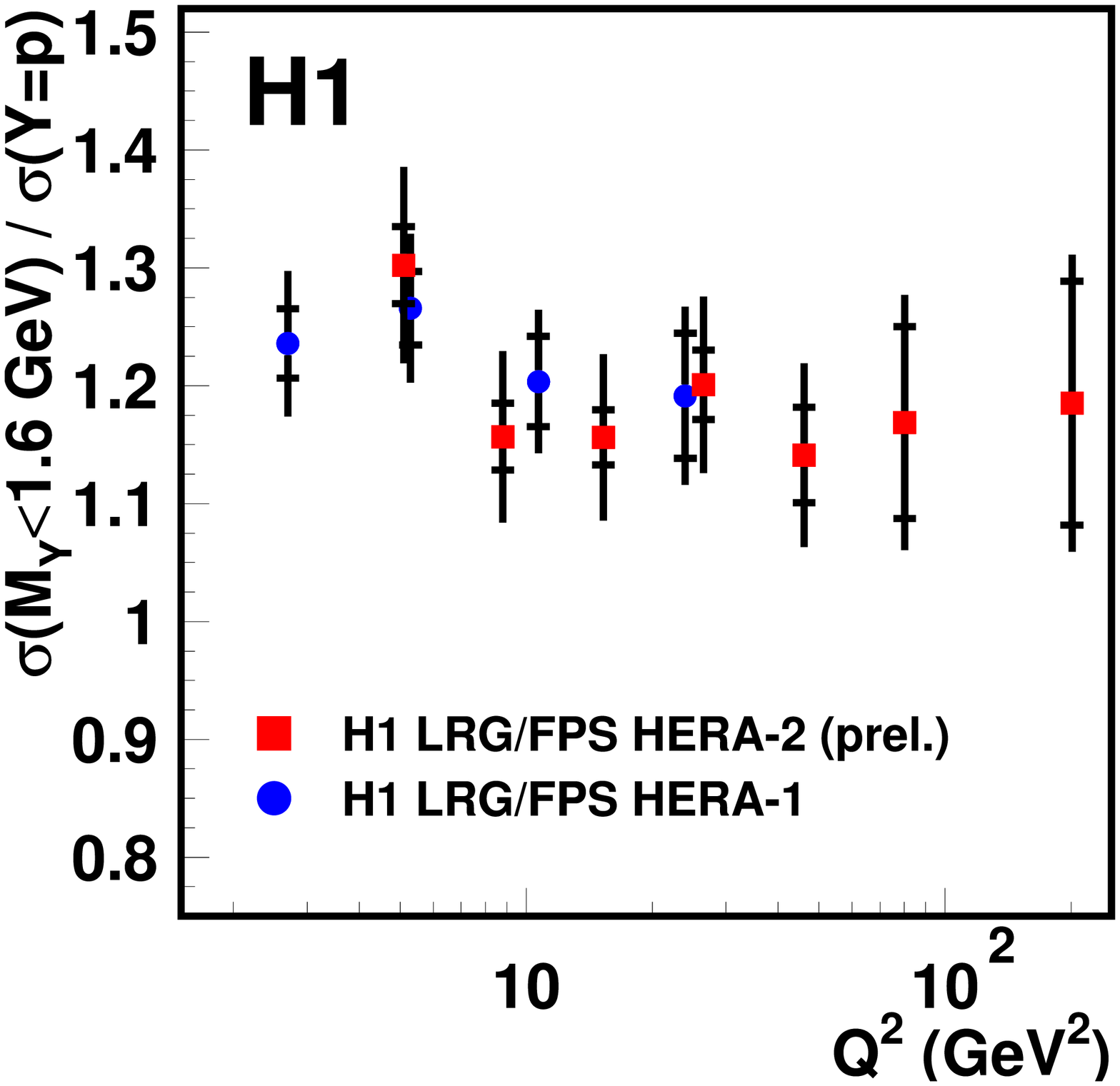}
\includegraphics[width=0.17\textwidth]%
{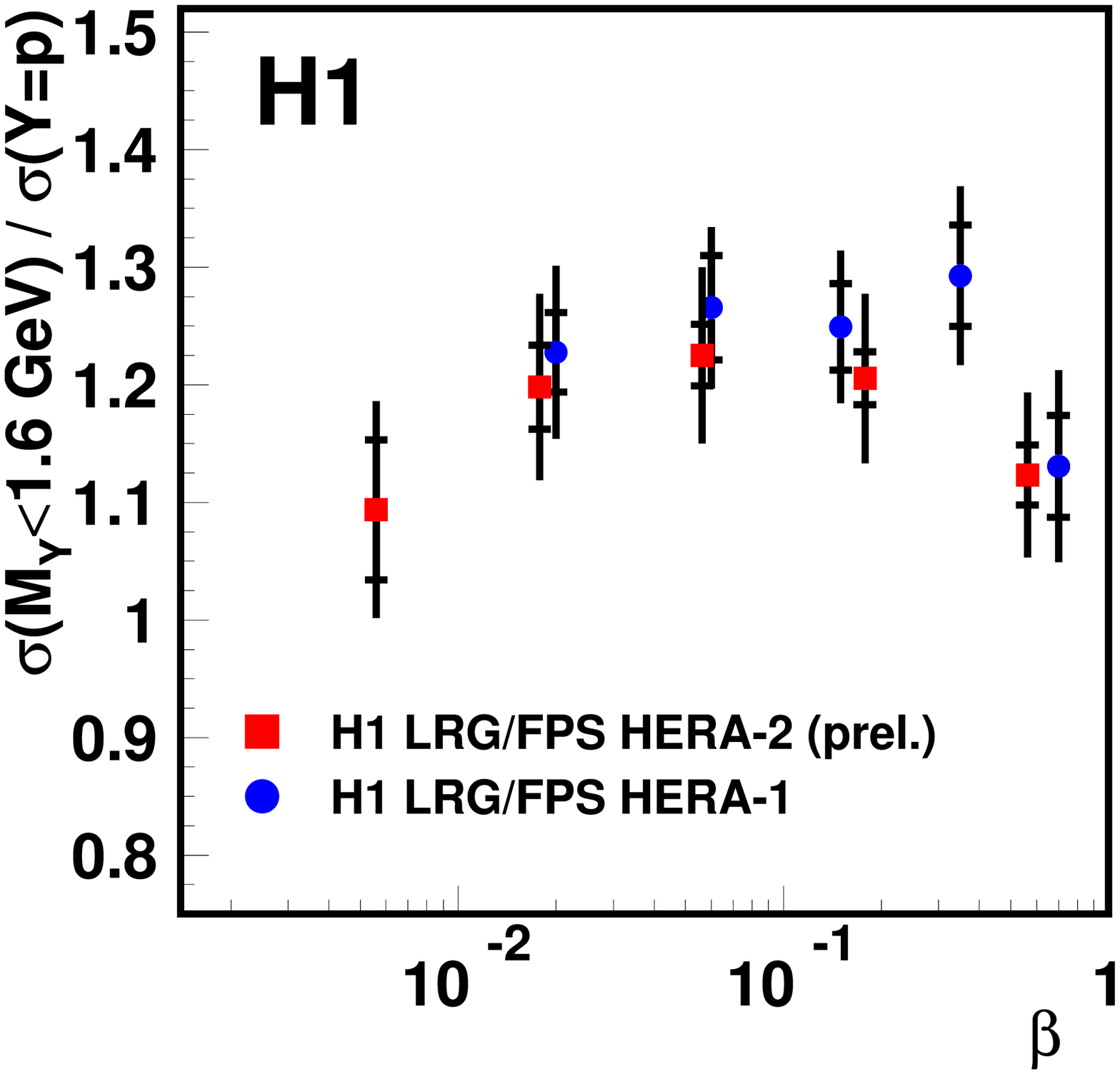}
\includegraphics[width=0.17\textwidth]%
{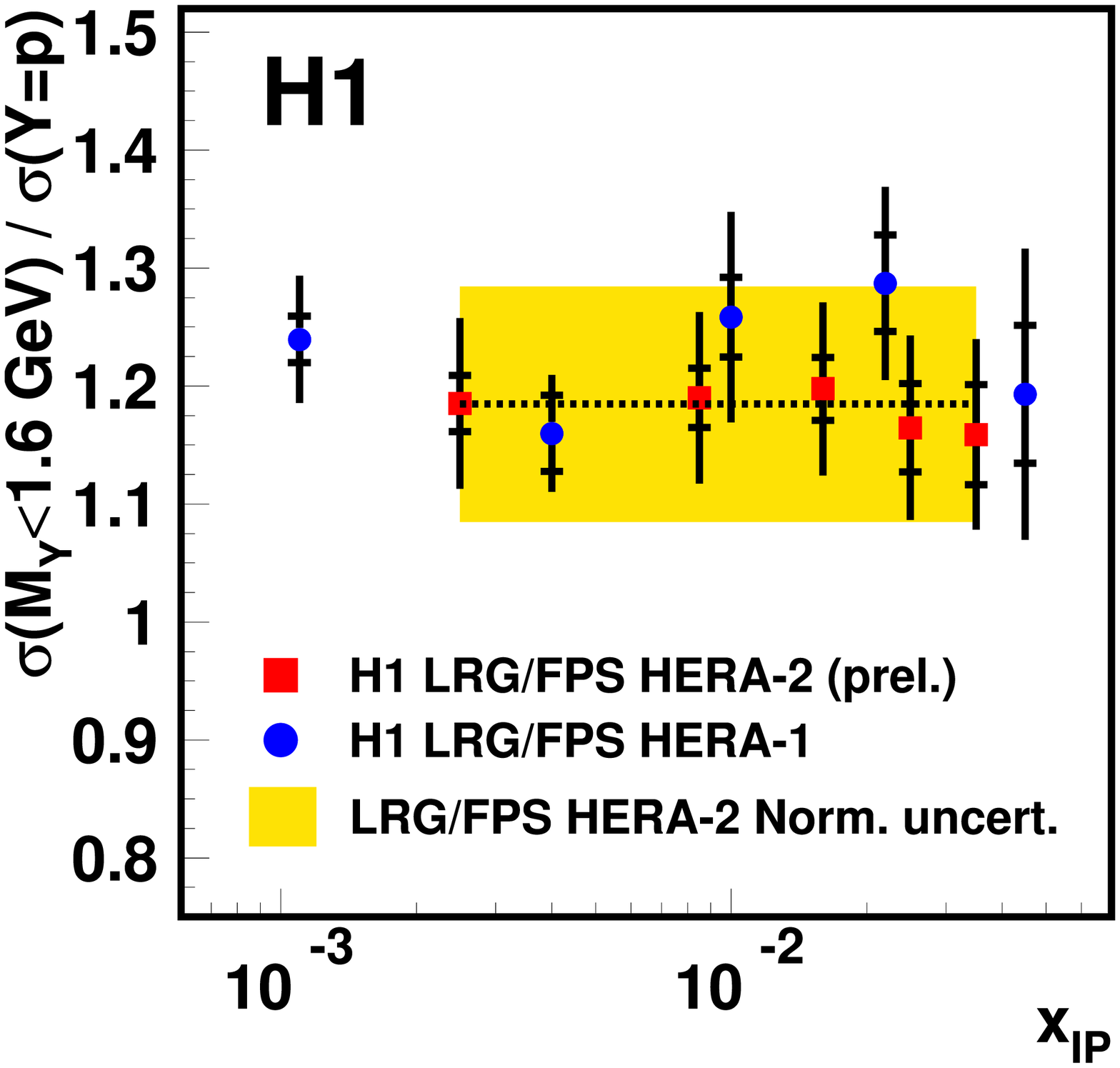}
}
\centerline{%
\includegraphics[width=0.43\textwidth]%
{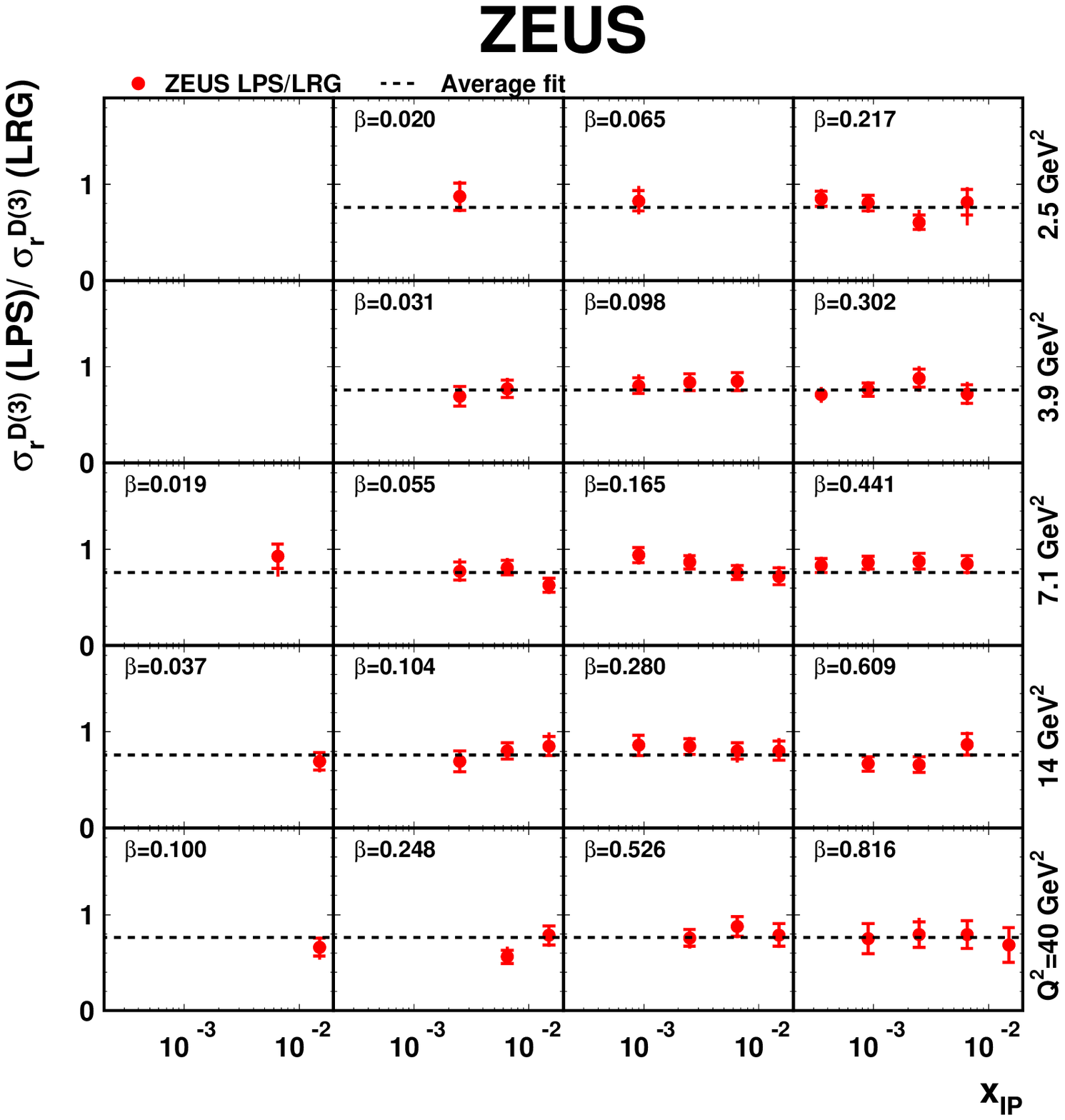}}
\vspace{-0.3cm}
\caption{%
The ratio of the reduced diffractive cross sections, 
$\sigma^{D(3)}_r$, obtained with the LPS and the LRG methods, as 
functions of $Q^2$, $\beta$ or $x_{\pomsub}$ (up, H1) or a function of 
$x_{\pomsub}$ for different ($Q^2$, $\beta$) bins (down, ZEUS). 
The proton dissociative background was not subtracted from the LRG data. 
The lines indicate the average value of the ratio. 
        }
\label{fig:csr3-lpsvslrgzeus} 
\end{figure}

To test the compatibility between the results from 
the different measurement techniques and to estimate the proton
dissociation admixture to the LRG data, 
the ratio of the $\sigma^{D(3)}_r$ values obtained with the PS method 
to those obtained with the LRG one was studied 
(Fig.~\ref{fig:csr3-lpsvslrgzeus}). 
No subtraction of the proton-dissociative admixture to the LRG data 
was performed. Within experimental uncertainties the ratio
is independent of $x_{\pomsub}$, $Q^2$ or $\beta$. 
The normalisation difference is ascribed to the proton dissociative 
contribution  which typically amounts $\sim 20\%$ of the LRG sample as 
estimated in both experiments. $\sigma^{D(3)}_r$ measurements with the 
PS and the LRG methods are consistent in the region of overlap. 

\section{Diffractive parton distribution functions}
\nin
The latest high precision set of DPDFs \cite{ZEUS-sigrdlrglps} was 
determined in the global QCD fits to ZEUS 
LRG and LPS cross section data \cite{ZEUS-DPDFsSJ2009} combined with 
data on diffractive dijet production in DIS \cite{ZEUS-djjdis}. 
Only data with $Q^2 >5$ GeV$^2$ could be fitted within the combined 
framework of DGLAP evolution and proton vertex factorisation. The

\vspace*{-0.05cm}
\nin
extracted DPDFs correspond to the single-diffractive reaction with a 
proton in the final state and are valid in the region 
$\vert t \vert <1$ GeV$^2$, $M_X >2$ GeV, $x_{\pomsub} < 0.1$.
The conventional DGLAP formalism in next-to-leading order (NLO) of QCD 
was applied. The parton distributi-

\vspace{-0.1cm}
\nin
ons at the starting scale ($Q^2_0$) 
were parameterised with polynomials in $z$:  $A z^{B}(1-z)^{C}$.
There were tried two fits (S and C) to the inclusive data  
and a fit (SJ) to  the combined inclusive and dijet data. 
Parameterisations of the gluon density at the starting 
scale  were taken to be very different to estimate theoretical 
uncertainties. Fits S and SJ were based on 
the parameterisation with $A_g, B_g$ and $C_g$ parameters free. 
In fit C the gluon density was taken to be a constant: 
                                        $A_g=const, B_g=0$ and $C_g=0$.   

\begin{figure}[!hbt] 
  \centerline{\includegraphics[width=0.48\textwidth]%
    {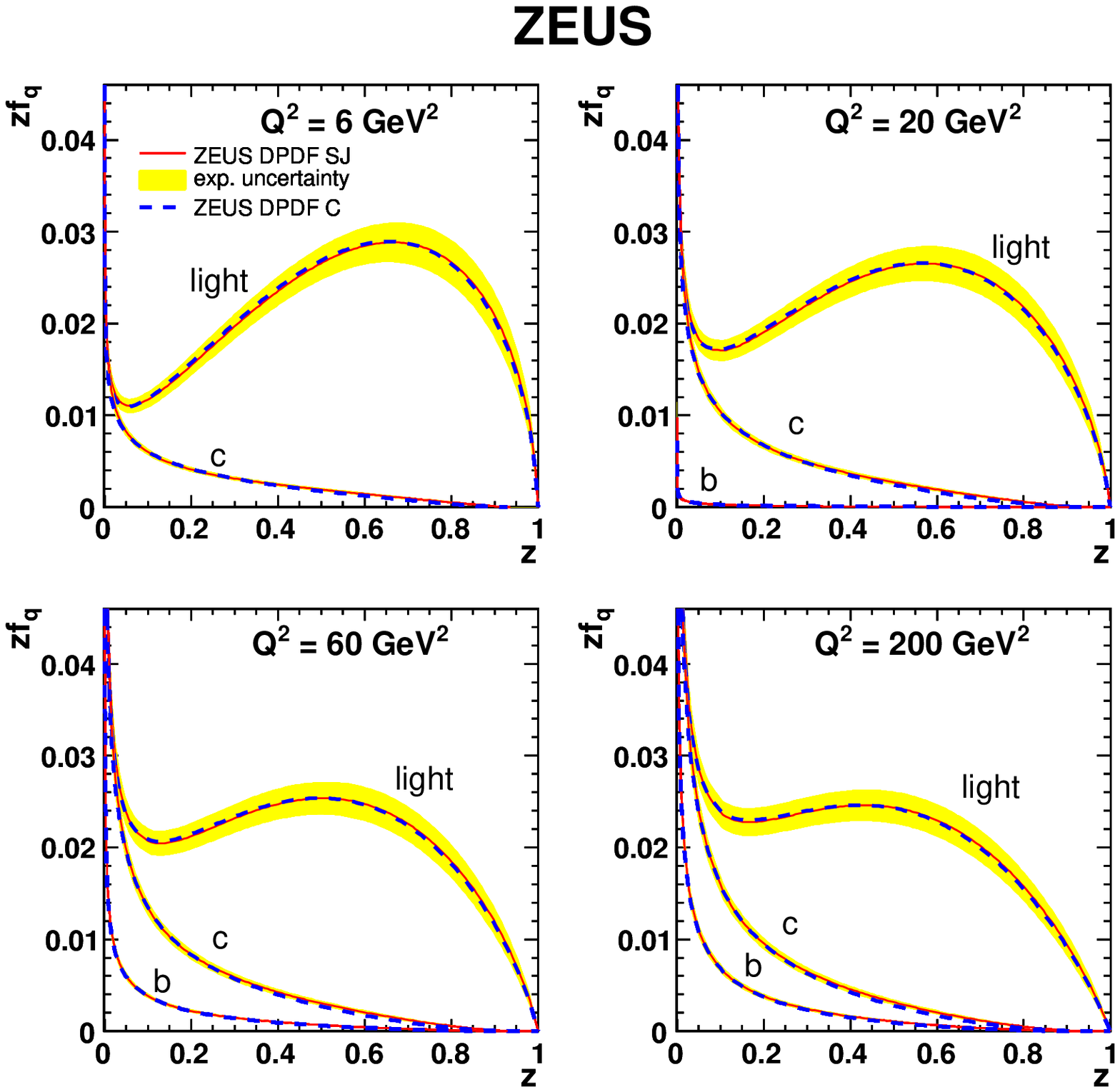}}
  \centerline{\includegraphics[width=0.48\textwidth]%
    {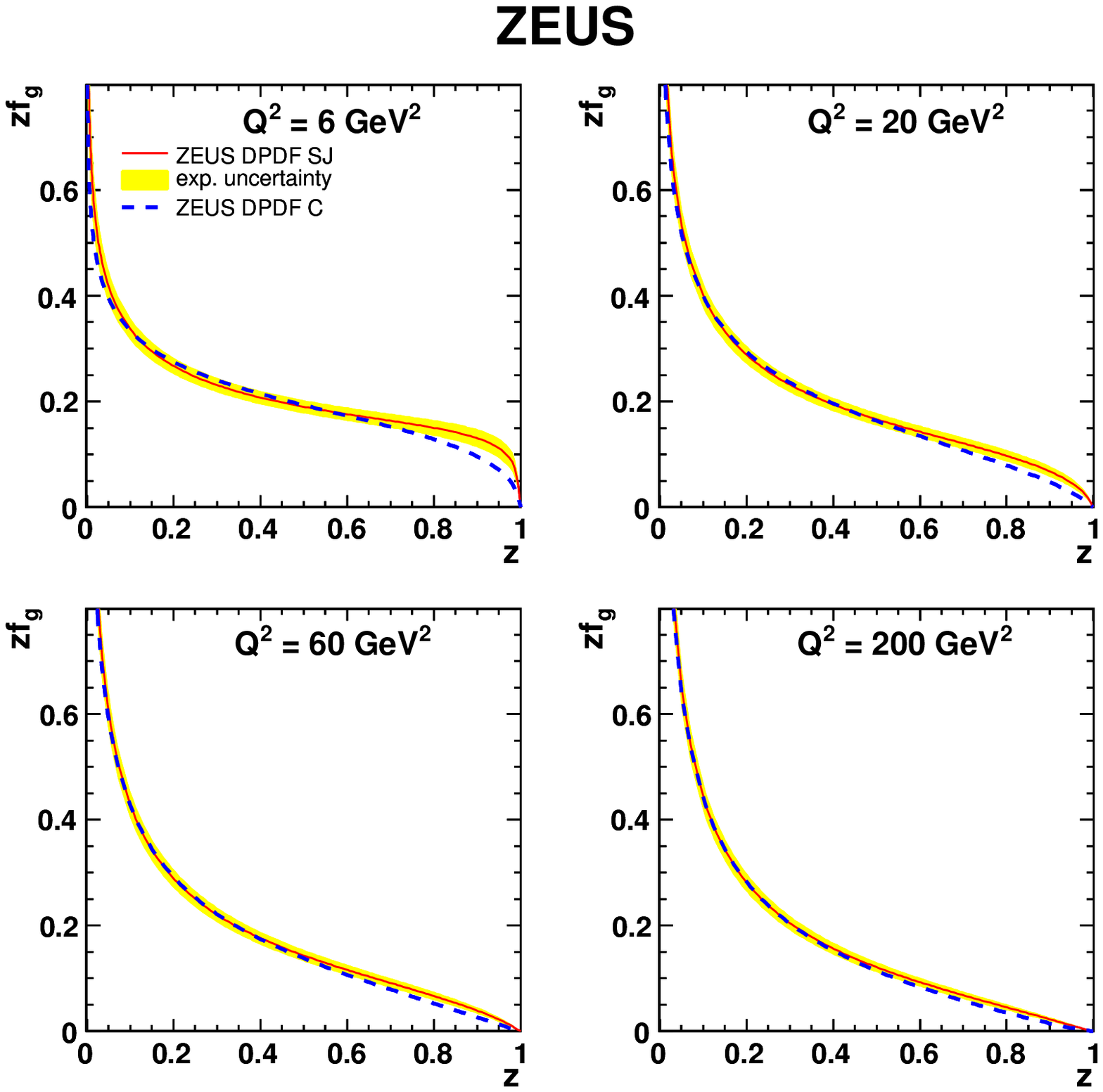}}
\vspace{-0.3cm}
\caption{%
Quark (up) and gluon (down) distributions from 
$\sigma_r^D(Q^2)$ measurements 
 with LRG, LPS and dijet ZEUS data. Shown are the total 
experimental uncertainties only. The starting scale was 
$Q^2_0= 1.8$ GeV$^2$ with strong coupling 
constant, $\alpha_s(M_Z) = 0.118$ and heavy quark masses 
$m_c=1.35$ GeV, $m_b=4.3$ GeV were used.
        }
\label{fig:dpdf-zqzg.2009.zeus} 
\end{figure}

The resulting quark and gluon densities are presented in 
Fig.~\ref{fig:dpdf-zqzg.2009.zeus}.
All fits (fit S is not shown) gave similar  distributions (upper plots) 
with small uncertainties for quark densities.
Fits S and C gave very different distributions for the  gluon density, 
particularly at high  $z$. Fit S grows rapidly with  $z\to 1$, while fit 
C vanishes as $z\to1$ in a smooth way.
The gluon density from fit SJ (down plots)
is similar to that of fit C and fit S is ruled out.  Diffractive dijet 
data are able to discriminate between different gluon parameterisations 
due to dominance of boson-gluon fusion mechanism in the production of 
jets. Combined fits to inclusive and dijet data constrain both the quark 
and gluon DPDFs to a good comparable precision across the whole z range.

 According to QCD factorisation, the DPDFs measured should be able to  
describe  cross sections for other diffractive processes: charm 
production in deep inelastic scattering \cite{ZEUS-F2cc} and jet 
photoproduction \cite{ZEUS-djjphp}. 
Predictions based on the fit ZEUS DPDF SJ 
 for the charm contribution to the diffractive structure 
function  were found in fair agreement with the data (plots not shown)
 for $x_{\pomsub}$ of 0.004 and 0.02 and $Q^2$ of 4 and 25 GeV$^2$. 
The predic-

\vspace{0.1cm}
\vspace{-0.2cm}
\nin
tions based on the fit ZEUS DPDF SJ for the 
diffractive dijet photoproduction cross sections in $x_{\gamma}^{obs}$
 , the fraction of the photon energy participated 
in producing the dijet system, and in the transverse energy $E_T^{jet}$
of the leading jet agree with  the data over the whole 
$x_{\gamma}^{obs}$ and $E_T^{jet}$ ranges (plots not shown). 
No suppression either of the resolved photon component, or of both 
components globally is observed. 

\section{Measurement of $F_L^D$}
\nin
The DIS diffractive reduced cross section at low $Q^2$  is determined by 
structure functions $F_2^D$ and $F_L^D$ in a linear form 
(see sec. 1). Thus measurements of $\sigma_r^D$  in fixed
 $(x,Q^2)$ bins at different $y$ allow to measure  
$F_2^D$ and $F_L^D$ simultaneously and in model independent manner. From 
linear fits to $\sigma_r^D$ in $y/Y_+$ one obtains $F_2^D$ as an intercept 
with the cross section axis and  $F_L^D$ being the slope of the fit. 
The precision of the fit depends on the range available in $y$.
Whilst  $F_2^D$ is determined by 
the sum of weighted quark and anti-quark distributions only, $F_L^D$ 
additionally includes gluon contribution which at low $x$ exceeds quark 
ones. Therefore  $F_L^D$ is a direct measure of the gluon distribution 
at low $x$ . Such measurement can be used for testing other methods.

\begin{figure}[hbt] 
  \centerline{\includegraphics[width=0.4\textwidth]
    {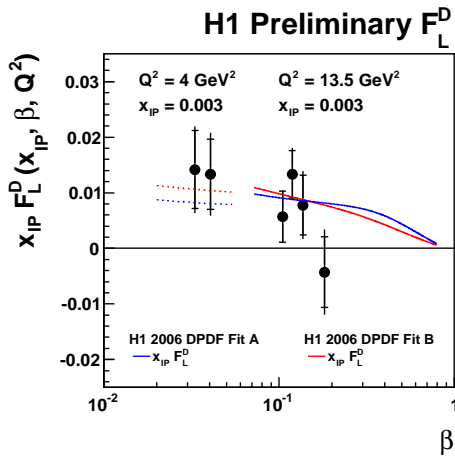}}
\vspace{-0.3cm}
\caption{
 $x_{\pomsub}F_L^D$ as a function of $\beta$ measured at 
$Q^2=4$ GeV$^2$ and $Q^2=13.5$ GeV$^2$ (earlier result). Predictions 
from NLO QCD  fits H1 2006 Fit A (blue line) and Fit B (red line) were 
estimated with starting scale $Q^2=8.5$ GeV$^2$ and extrapolated to 
$Q^2=4$ GeV$^2$ (dotted lines)
       }
\label{fig:h1fldprel.2009} 
\end{figure}

As $Q^2=sxy$ different values of $y$ can be reached by variation in $s$. 
 $F_L^D(x,Q^2)$ was measured for the first time 
by the H1 experiment \cite{H1FLD-DIS2010} 
in the range $2.5<Q^2<7$ GeV$^2$ and 
$0.001<x_{\pomsub}<0.01$ at proton beam energies 460, 575 and 920 GeV. 
Within the experimental uncertainties
the preliminary result of  the measurements 
(Fig.~\ref{fig:h1fldprel.2009}).  is consistent with the predictions 
based on H1 2006 DPDF \cite{H1-DPDFs2006} Fit A and Fit B extrapolated 
to low $Q^2$, i.e.
being in good agreement with the NLO QCD picture of diffraction.

\vspace{-0.3cm}
\vspace{-0.2cm}
\nin

\section{Conclusions}
\label{sec:conc}
\nin 
  The full HERA data sample has been analysed 
to measure diffractive parton density functions 
 to the best precision possible.
As  it is desirable to combine all available diffractive DIS data sets 
into an unique  HERA set, H1 and ZEUS measurements by various methods 
with very different systematics have been compared in detail to test the 
control over the systematics and  agreement between the different 
measurements. 

All measurements are found to be consistent in the shapes of the 
distributions throughout most of the phase space.  
Compatible estimations of the proton dissociation contributions in  the 
LRG samples were obtained. There was found a 
global normalisation difference at the $13\%$ level between the similar
measurements of the two experiments. 
The data are well described by the QCD fit and the quark densities (from
the inclusive data) and gluon densities (from the dijet data) are 
constrained to similar precision. 
The first $F^D_L$ measurement is in agreement with DPDF prediction. 

Since H1 and ZEUS diffractive data are not yet combined, further 
improvement in accuracy of DPDFs is possible.
%
%

\section*{Acknowledgements}
\nin
It is a pleasure to thank  DESY  for providing 
me with  financial support.
%













%

\end{document}